# Generalized confluent hypergeometric solutions of the Heun confluent equation

T.A. Ishkhanyan[1,2] and A.M. Ishkhanyan[1,3]

[1]Russian-Armenian University, Yerevan, 0051 Armenia
[2]Institute for Physical Research, NAS of Armenia, Ashtarak, 0203 Armenia
[3]Institute of Physics and Technology, National Research Tomsk Polytechnic University, Tomsk, 634050 Russian Federation

We show that the Heun confluent equation admits infinitely many solutions in terms of the confluent generalized hypergeometric functions. For each of these solutions a characteristic exponent of a regular singularity of the Heun confluent equation is a non-zero integer and the accessory parameter obeys a polynomial equation. Each of the solutions can be written as a linear combination with constant coefficients of a finite number of either the Kummer confluent hypergeometric functions or the Bessel functions.

**PACS numbers:** 02.30.Gp Special functions, 02.30.Hq Ordinary differential equations, 02.30.Mv Approximations and expansions

**MSC numbers**: 33E30 Other functions coming from differential, difference and integral equations, 34B30 Special equations (Mathieu, Hill, Bessel, etc.), 30Bxx Series expansions

**Keywords:** Confluent Heun equation, confluent hypergeometric function, Bessel function, recurrence relation

## 1. Introduction

The Heun confluent equation [1-3] is a second order linear differential equation widely encountered in contemporary physics research ranging from hydrodynamics, polymer and chemical physics to atomic and particle physics, theory of black holes, general relativity and cosmology, etc. (see, e.g., [4-22] and references therein). This equation has two regular singularities conventionally located at points $z=0$ and $z=1$ of complex $z$-plane, and an irregular singularity of rank 1 at $z=\infty$. Due to such a specific structure of singularities, the Heun confluent equation presents a generalization of both the Gauss ordinary and the Kummer confluent hypergeometric equations widely applied in physics during the past century. We adopt here the following canonical form of the Heun confluent equation [3]:

$$\frac{d^2u}{dz^2}+\left(\frac{\gamma}{z}+\frac{\delta}{z-1}+\varepsilon\right)\frac{du}{dz}+\frac{\alpha z-q}{z(z-1)}u=0, \qquad (1)$$

from which both hypergeometric equations are obtained by simple choices of the parameters. Another prominent equation that presents a particular case of this equation is the algebraic form of the Mathieu equation which is obtained if $\varepsilon=0$ and $\gamma=\delta=1/2$ [1-3].

Despite the considerable research devoted to the mathematical properties of equation (1), it is still much less studied than its hypergeometric predecessors or the Mathieu equation, and the solutions in terms of simpler functions, including the special functions of the hypergeometric class, are rare. In this brief communication we introduce infinitely many solutions in terms of generalized hypergeometric functions [23,24]. The result is that for the general case $\varepsilon \neq 0$ there exist infinitely many solutions in terms of a single generalized hypergeometric function ${}_pF_p$, while for the reduced case $\varepsilon = 0$ there are infinitely many solutions in terms of a single function ${}_pF_{p+1}$. In both cases of non-zero or zero $\varepsilon$ the solutions exist if a characteristic exponent of a regular singularity of the Heun confluent equation is a non-zero integer and the accessory parameter $q$ obeys a polynomial equation.

## 2. Solutions for non-zero $\varepsilon$

Let $\varepsilon \neq 0$. The characteristic exponents of the singularity $z = 1$ are $\mu_{1,2} = 0, 1-\delta$. Let the exponent $\mu_2 = 1-\delta$ is a non-zero integer. A basic observation is that for any *negative* integer $\delta = -N$, $N = 1, 2, 3, ...$ (the case of a positive integer $\delta$ is discussed afterwards) the Heun confluent equation admits a solution given as

$$u = {}_{N+1}F_{1+N}(1+e_1,...,1+e_N,\alpha/\varepsilon;e_1,...,e_N,\gamma;-\varepsilon z). \tag{2}$$

This solution applies for certain particular choices of the accessory parameter $q$ defined by a polynomial equation of the degree $N+1$. We note that for $\delta = 0$ the Heun confluent equation admits a solution in terms of the Kummer confluent hypergeometric function:

$$u = {}_1F_1(\alpha/\varepsilon;\gamma;-\varepsilon z), \tag{3}$$

achieved for 
$$q - \alpha = 0. \tag{4}$$

The solution for $N = 1$ reads

$$u = {}_2F_2(\alpha/\varepsilon, 1+e_1;\gamma, e_1;-\varepsilon z), \quad \delta = -1, \tag{5}$$

$$q^2 - (2\alpha + \gamma - 1 + \varepsilon)q + \alpha(\alpha + \gamma + \varepsilon) = 0, \tag{6}$$

where the parameter $e_1$ is given as $e_1 = \alpha/(q-\alpha)$. This solution was noticed by Letessier [25,26] and studied by Letessier, Valent and Wimp [27]. We note that the parameter $e_1$ parameterizes the root of equation (6) as

$$q = \alpha \frac{1+e_1}{e_1}, \quad 1 = \frac{e_1(1+e_1-\gamma)}{\varepsilon e_1 - \alpha}. \tag{7}$$

The solution for $N = 2$ is

$$u = {}_3F_3(\alpha/\varepsilon, 1+e_1, 1+e_2;\gamma, e_1, e_2;-\varepsilon z), \quad \delta = -2, \tag{8}$$

$$2(q-\alpha)(\alpha+\varepsilon) + (q-\alpha-2(\gamma-1+\varepsilon))(q^2-(2\alpha+\gamma-2+\varepsilon)q+\alpha(\alpha+\gamma+\varepsilon)) = 0, \tag{9}$$

where the parameters $e_{1,2}$ are defined by the equations

$$q = \alpha \frac{(1+e_1)(1+e_2)}{e_1 e_2}, \quad 1 = \frac{e_1(1+e_1-\gamma)}{(\varepsilon e_1 - \alpha)} \frac{e_2(1+e_2-\gamma)}{(\varepsilon e_2 - \alpha)}, \tag{10}$$

and the solution for $N = 3$ reads

$$u = {}_4F_4(\alpha/\varepsilon, 1+e_1, 1+e_2, 1+e_3; \gamma, e_1, e_2, e_3; -\varepsilon z), \quad \delta = -3, \tag{11}$$

$$3(\alpha+2\varepsilon)\left(q^2 - (2\alpha+\gamma-3+\varepsilon)q + \alpha(\alpha+\gamma+\varepsilon)\right) + \left(q-\alpha-3(\gamma-1+\varepsilon)\right) \times$$
$$\left(4(q-\alpha)(\alpha+\varepsilon) + \left(q-\alpha-2(\gamma-2+\varepsilon)\right)\left(q^2 - (2\alpha+\gamma-3+\varepsilon)q + \alpha(\alpha+\gamma+\varepsilon)\right)\right) = 0 \tag{12}$$

where the parameters $e_{1,2,3}$ obey the equations

$$q = \alpha \prod_{k=1}^{3} \frac{1+e_k}{e_k}, \quad 1 = \prod_{k=1}^{3} \frac{e_k(1-\gamma+e_k)}{(\varepsilon e_k - \alpha)}, \tag{13}$$

and

$$q = \alpha - \sum_{n=1}^{3}(e_n + n - \gamma - \varepsilon). \tag{14}$$

(One should be careful that not all solutions of equations (13),(14) give the correct values of $e_{1,2,3}$; for the unique determination of $e_{1,2,3}$ see below.)

In the general case $\delta = -N$ the accessory parameter $q$ and the parameters $e_{1,2,\ldots,N}$ involved in solution (2) are determined from a system of $N+1$ algebraic equations. These equations are constructed by equating to zero the coefficients of the following polynomial $\Pi(n)$ in an auxiliary variable $n$:

$$\Pi = (\alpha + \varepsilon(n-1))\prod_{k=1}^{N}(e_k + n) + Q\prod_{k=1}^{N}(e_k - 1 + n) - (n-1)(\gamma - 2 + n)\prod_{k=1}^{N}(e_k - 2 + n), \tag{15}$$

where

$$Q = -q + (n-1)(n-2+\gamma+\delta-\varepsilon). \tag{16}$$

An important point is that the polynomial $\Pi(n)$ is of degree $N$, not $N+2$, as it may be supposed at first glance. This is because the two possible highest-degree terms proportional to $n^{N+1}$ and $n^{N+2}$ actually vanish. We thus have $N+1$ equations, of which $N$ equations serve for determination of the parameters $e_{1,2,\ldots,N}$ and the remaining one, after elimination of $e_{1,2,\ldots,N}$, imposes a restriction on the parameters of the Heun confluent equation. This restriction is checked to be a polynomial equation of the degree $N+1$ for the accessory parameter $q$. Examples of this equation for $N = 0,1,2,3$ are those given by equations (4), (6),(9), and (12). A concluding remark is that the system of the algebraic equations at hand leads to the following generalization of equations (13),(14):

$$q = \alpha \prod_{k=1}^{N} \frac{1+e_k}{e_k}, \quad 1 = \prod_{k=1}^{N} \frac{e_k(1-\gamma+e_k)}{(\varepsilon e_k - \alpha)}, \quad q = \alpha - \sum_{n=1}^{N} (e_n + n - \gamma - \varepsilon). \tag{17}$$

The derivation of the presented results is discussed in the next section.

## 3. Derivations for the case $\varepsilon \neq 0$

Consider the Frobenius series solution of the Heun confluent equation (1) for the vicinity of the singularity $z = 0$:

$$u = z^{\mu} \sum_{n=0}^{\infty} c_n z^n, \quad \mu = 0, 1-\gamma. \tag{18}$$

The coefficients of this expansion obey a three-term recurrence relation:

$$R_n c_n + Q_{n-1} c_{n-1} + P_{n-2} c_{n-2} = 0. \tag{19}$$

For the exponent $\mu = 0$ the coefficients of this relation read

$$R_n = (\gamma - 1 + n)n, \tag{20}$$

$$Q_n = q - (\gamma + \delta - \varepsilon - 1 + n)n, \tag{21}$$

$$P_n = -(\alpha + \varepsilon n). \tag{22}$$

The idea is to look for the cases when the Frobenius expansion (18) is reduced to a confluent generalized hypergeometric series. To examine this possibility, we note that the generalized hypergeometric function ${}_pF_p$ of a scaled argument $s_0 z$ is defined through the power series [23,24]

$${}_pF_p(a_1,...,a_p; b_1,...,b_p; s_0 z) = \sum_{n=0}^{\infty} c_n (s_0 z)^n \tag{23}$$

the coefficients of which obey the two-term recurrence relation

$$\frac{c_n}{c_{n-1}} = \frac{s_0}{n} \frac{\prod_{k=1}^{p}(a_k - 1 + n)}{\prod_{k=1}^{p}(b_k - 1 + n)}. \tag{24}$$

Having in the mind the function (2), we put $p = N + 1$ and

$$a_1,...,a_N, a_{N+1} = 1+e_1,...,1+e_N, a, \tag{25}$$

$$b_1,...,b_N, b_{N+1} = e_1,...,e_N, b. \tag{26}$$

The recurrence relation (24) is then rewritten as

$$\frac{c_n}{c_{n-1}} = \frac{(a-1+n)s_0}{(b-1+n)n} \prod_{k=1}^{N} \frac{e_k + n}{e_k - 1 + n}. \tag{27}$$

Substituting this into equation (19), we have

$$R_n \frac{(a-1+n)s_0}{(b-1+n)n} \prod_{k=1}^{N} \frac{e_k+n}{e_k-1+n} + Q_{n-1} + P_{n-2} \frac{(b-2+n)(n-1)}{(a-2+n)s_0} \prod_{k=1}^{N} \frac{e_k-2+n}{e_k-1+n} = 0. \quad (28)$$

$R_n$ cancels the factor $(b-1+n)n$ of the denominator of the first term of this equation if we choose $b = \gamma$. Similarly, $P_{n-2}$ cancels the factor $(a-2+n)$ of the denominator of the last term if $a = \alpha/\varepsilon$. Hence, we put $b = \gamma$, $a = \alpha/\varepsilon$. Then, cancelling the common denominator, equation (28) is rewritten as

$$s_0 \left(\frac{\alpha}{\varepsilon} - 1 + n\right) \prod_{k=1}^{N}(e_k+n) + Q_{n-1} \prod_{k=1}^{N}(e_k-1+n) - \frac{\varepsilon}{s_0}(n-1)(\gamma-2+n) \prod_{k=1}^{N}(e_k-2+n) = 0. \quad (29)$$

This is a polynomial equation in $n$ of the degree $N+2$. The coefficient of the highest-degree term $\sim n^{N+2}$ is $-(1+\varepsilon/s_0)$. Hence, for $s_0 = -\varepsilon$ this term vanishes and equation (29) takes the form

$$\sum_{m=0}^{N+1} A_m(a,q;\alpha,\beta,\gamma,\delta,\varepsilon;e_1,...,e_N)n^m = 0. \quad (30)$$

Now, equating to zero the coefficients $A_m$ warrants the satisfaction of the recurrence relation (19) for all $n$. We thus have $N+2$ equations $A_m = 0$, $m = 0,1,..,N+1$, of which $N$ equations serve for determination of the parameters $e_{1,2,...,N}$ and the remaining two impose restrictions on the parameters of the Heun confluent equation (1).

One of these restrictions is readily derived by calculating the coefficient $A_{N+1}$ of the term proportional to $n^{N+1}$, which is shown to be $N+\delta$. Hence,

$$\delta = -N. \quad (31)$$

The second restriction is derived by elimination of $e_{1,2,...,N}$. For $N = 0,1,2,3$ these restrictions are those given by equations (4),(6),(9), and (12), respectively. For higher $N$ the equations are cumbersome; we omit those. We note, however, that this restriction can alternatively be derived via termination of the series solution of the Heun confluent equation in terms of the Kummer confluent hypergeometric functions [28]. This assertion is deduced if we recall that the generalized hypergeometric function (2), with $N$ numerator parameters exceeding the denominator ones by unity, can be written as a linear combination with constant coefficients of a finite number of the confluent hypergeometric functions. This linear combination can conveniently be derived by the termination of the expansions of the solutions of the Heun confluent equation in terms of the Kummer functions [28-31]. The termination condition for $\delta = -N$ is a polynomial equation of degree $N+1$ for the accessory parameter $q$.

As regards equations (17), the first one is derived from equation (29) by putting $n = 1$, the third equation comes from the coefficient of the term proportional to $n^N$, and the second equation is a result of numerical simulations that we have carried out for $N$ up to 20. Though we have checked the validity of this equation analytically for $N \leq 7$, we have not a proof for arbitrary $N$. However, this equation is just an additional observation. It is not necessary for construction of the solution of the Heun confluent equation. The $N+1$ equations $A_m = 0$ suffice. This fulfils the development. A complementary remark is that the solution of the system $A_m = 0$ is unique up to the transposition of the parameters $e_{1,2,\ldots,N}$.

**4. Positive integer $\delta \neq 1$ or arbitrary integer $\gamma \neq 1$**

Let $\delta$ is now a *positive* integer: $\delta = N$, $N = 1, 2, 3, \ldots$. Solutions for this case are constructed by applying the elementary power change $u = (z-1)^{1-\delta} w$ which transforms the Heun confluent equation into another Heun confluent equation with the altered parameter $\delta_1 = 2 - \delta$. For $\delta \geq 2$ we get a Heun confluent equation with a zero or negative integer $\delta_1$. As a result, we derive the solution

$$u = (z-1)^{1-\delta} {}_{N+1}F_{1+N}\left(\frac{\alpha}{\varepsilon} + 1 - \delta, \tilde{e}_1 + 1, \ldots, \tilde{e}_N + 1; \gamma, \tilde{e}_1, \ldots, \tilde{e}_N; -\varepsilon z\right). \tag{32}$$

Thus, the only exception is the case $\delta = 1$ for which both characteristic exponents $\mu_{1,2} = 0, 1-\delta$ are zero. We do not know a ${}_pF_p$ solution for this exceptional case.

Solutions in terms of the generalized confluent hypergeometric functions of similar structure can also be constructed for any integer $\gamma = N \in \mathbb{Z}$, $N \neq 1$. These solutions are derived by employing the Frobenius expansion in the vicinity of the singular point $z = 1$. The solution for a negative integer $\gamma = -N$ is of the form

$$u = {}_{N+1}F_{1+N}\left(\frac{\alpha}{\varepsilon}, e_1 + 1, \ldots, e_N + 1; e_1, \ldots, e_N, \delta; -\varepsilon(z-1)\right), \tag{33}$$

and for a positive integer $\gamma \neq 1$ the solution is given as

$$u = z^{1-\gamma} {}_{N+1}F_{1+N}\left(\frac{\alpha}{\varepsilon} + 1 - \gamma, e_1 + 1, \ldots, e_N + 1; e_1, \ldots, e_N, \delta; -\varepsilon(z-1)\right). \tag{34}$$

We note that here instead of $\gamma$ we have $\delta$ as a denominator parameter. It is worth mentioning that for any choice of parameters the generalized hypergeometric series involved in all presented solutions converge everywhere in the complex $z$-plane (with the proviso that none of the denominator parameters is zero or a negative integer) [23,24].

## 5. Reduced case $\varepsilon = 0$

The case $\varepsilon = 0$ is special since the nature of the singularity at $z = \infty$ is changed to have an asymptote inferred from the *subnormal* Thomé solutions (for non-zero $\varepsilon$ the asymptote is given by the normal Thomé solutions) [32]. We note that this reduced case (which can be viewed as the Whittaker-Ince limit of the generalized spheroidal wave equation [33]) degenerates to the Gauss hypergeometric equation if $\alpha = 0$ and presents the algebraic form of the Mathieu equation if $\alpha \neq 0$, $\gamma = \delta = 1/2$.

The confluent generalized hypergeometric solutions of the Heun confluent equation in this case are as follows. Let $\delta$ is a *non-positive* integer. A simple case is the Bessel-function solution for $\delta = 0$:

$$u = (\alpha z)^{\frac{1-\gamma}{2}} J_{\gamma-1}(2\sqrt{\alpha z}) = {}_0F_1(;\gamma;-\alpha z), \tag{35}$$

valid for
$$q - \alpha = 0. \tag{36}$$

A basic result is now that for any $\delta = -N$, $N = 1, 2, 3, \ldots$ (as for the previous case $\varepsilon \neq 0$, the case of a positive integer $\delta$ is treated separately) the Heun confluent equation admits a solution given as

$$u = {}_NF_{1+N}(1+e_1,\ldots,1+e_N; e_1,\ldots,e_N,\gamma;-\alpha z). \tag{37}$$

This solution is deduced from solution (2) by a limiting procedure or by a slight modification of the derivation lines described in the previous section.

The solution for $N = 1$ reads

$$u = {}_1F_2(1+e_1; \gamma, e_1; -\alpha z), \quad \delta = -1, \tag{38}$$

$$q^2 - (2\alpha + \gamma - 1)q + \alpha(\alpha + \gamma) = 0, \tag{39}$$

where the parameter $e_1$ is given as $e_1 = \alpha/(q-\alpha)$.

For $N = 2$ we have

$$u = {}_2F_3(1+e_1, 1+e_2; \gamma, e_1, e_2; -\alpha z), \quad \delta = -2, \tag{40}$$

$$2(q-\alpha)\alpha + (2+q-\alpha-2\gamma)\left(q^2 - (2\alpha+\gamma-2)q + \alpha(\alpha+\gamma)\right) = 0, \tag{41}$$

where the parameters $e_{1,2}$ are defined by the equations

$$q = \alpha \frac{(1+e_1)(1+e_2)}{e_1 e_2}, \quad 1 = \frac{e_1(1+e_1-\gamma)}{\alpha}\frac{e_2(1+e_2-\gamma)}{\alpha}, \tag{42}$$

and the solution for $N = 3$ is

$$u = {}_3F_4(1+e_1, 1+e_2, 1+e_3; \gamma, e_1, e_2, e_3; -\alpha z), \quad \delta = -3, \tag{43}$$

$$3\alpha\left(q^2-(2\alpha+\gamma-3)q+\alpha(\alpha+\gamma)\right)+(q-\alpha-3(\gamma-1))\times$$
$$\left(4(q-\alpha)\alpha+(q-\alpha-2(\gamma-2))\left(q^2-(2\alpha+\gamma-3)q+\alpha(\alpha+\gamma)\right)\right)=0 \quad (44)$$

where the parameters $e_{1,2,3}$ obey the equations

$$q=\alpha\prod_{k=1}^{3}\frac{1+e_k}{e_k}, \quad 1=\prod_{k=1}^{3}\frac{e_k(1-\gamma+e_k)}{(-\alpha)}, \quad q=\alpha-\sum_{n=1}^{3}(e_n+n-\gamma). \quad (45)$$

In the general case $\delta=-N$ the accessory parameter $q$ and the parameters $e_{1,2,...,N}$ are determined from a system of $N+1$ algebraic equations. These equations are again derived by equating to zero the coefficients of the auxiliary polynomial $\Pi(n)$ given by equations (15),(16), where one should put $\varepsilon=0$.

We would like to conclude by noting that the solution for a positive integer $\delta \neq 1$ as well as the solution for an integer (negative or positive) $\gamma \neq 1$ are constructed exactly in the same manner as described above for the case of non-zero $\varepsilon$ (we recall that the latter solution is constructed by employing the Frobenius expansion near the regular singularity $z=1$).

## 6. Discussion

Thus, we have presented infinitely many solutions of the Heun confluent equation, each written in terms of a single generalized confluent hypergeometric function. This is the extension to the confluent case of our recent results for the Heun general equation [34]. The existence of the latter solutions (applicable for the Fuchsian differential equations which have only regular singularities) has been conjectured (and proved for the first five lower-order cases) by Takemura [35] based on the earlier results by Letessier et al. [27].

As it was mentioned in the introduction, the Heun confluent equation has a wide coverage in contemporary physics and mathematics, so that one may expect many applications of these solutions. Here is an example from quantum few-state non-adiabatic dynamics [36-38].

The semi-classical time-dependent two-state problem is written as a system of coupled first-order differential equations for probability amplitudes $a_1(t)$ and $a_2(t)$ of two states of a quantum system driven by a quasi-resonant external field with amplitude modulation $U(t)$ and phase modulation $\delta(t)$ [36]:

$$i\frac{da_1}{dt}=Ue^{-i\delta}a_2, \quad i\frac{da_2}{dt}=Ue^{+i\delta}a_1. \quad (46)$$

This system is equivalent to the second-order ordinary differential equation

$$\ddot{a}_2 + \left(-i\dot{\delta} - \frac{\dot{U}}{U}\right)\dot{a}_2 + U^2 a_2 = 0, \tag{47}$$

where the over-dots denote differentiation with respect to time.

Consider the excitation by optical laser radiation with the field-configuration given as

$$U = U_0, \quad \delta_t = \delta_0 - \frac{\delta_0 + \delta_1}{1 + W(e^{-t/\tau})}, \tag{48}$$

where $U_0, \delta_0, \delta_1$ are arbitrary constants and $W$ is the Lambert-W function [39,40]. This is a constant-amplitude field-configuration describing asymmetric crossing of the resonance at a time point $t/\tau = -\delta_1/\delta_0 - \ln(\delta_1/\delta_0)$. This field configuration is a member of the confluent Heun two-state models presented in [10] (class $k_{1,2} = (-1,+1)$, Eq. (49) of [10]). It has been shown that the two-state problem (47) for this model is reduced, by applying the transformation $\psi = z^{\alpha_1} e^{\alpha_0 z} u(z)$, $z = -W(e^{-t/\tau})$, to the Heun confluent equation with $\delta = -1$. It can further be checked that the parameters of the latter equation are such that they satisfy equation (6) for the accessory parameter $q$. With this, using equation (5), we obtain a fundamental solution of the two-state problem in terms of the Goursat generalized hypergeometric function $_2F_2$ [41] explicitly written as

$$a_2 = z^{\alpha_1} e^{\alpha_0 z} \,_2F_2\left(a, 1+e; e, \gamma; \varepsilon W(e^{-t/\tau})\right), \tag{49}$$

where
$$\gamma = 1 + 2\alpha_1 - i\tau\delta_1, \quad \varepsilon = 2\alpha_0 - i\tau\delta_0,$$

$$a = \frac{\delta_0 \delta_1 - 4U_0^2}{2\varepsilon/\tau^2} + \frac{\gamma - 1}{2}, \quad e = \alpha_1 - i\tau\delta_1 + \alpha_0 - i\tau\delta_0, \tag{50}$$

$$\alpha_1 = \frac{i\tau}{2}\left(\delta_1 \pm \sqrt{\delta_1^2 + 4U_0^2}\right), \quad \alpha_0 = \frac{i\tau}{2}\left(\delta_0 \pm \sqrt{\delta_0^2 + 4U_0^2}\right). \tag{51}$$

Here any combination of signs of the roots involved in $\alpha_1$ and $\alpha_0$ is applicable. This can be used to construct the general solution of the problem.

We conclude by a complementary observation that equation (47) can readily be changed into the Schrödinger form with missing first-derivative term. As a result, one then arrives at a generalized confluent hypergeometric representation of the solution of the Schrödinger problem for the Lambert-W step potential for which the solution was initially written as a linear combination of two Kummer confluent hypergeometric functions [42]. Many more such representations can immediately be constructed if other wave equations (both relativistic and non-relativistic) as well as other solvable few-state models are applied.

**Acknowledgments**

The work has been supported by the Armenian State Committee of Science (SCS Grant No. 18RF-139), the Armenian National Science and Education Fund (ANSEF Grant No. PS-4986), the Russian-Armenian (Slavonic) University at the expense of the Ministry of Education and Science of the Russian Federation, and the project "Leading Russian Research Universities" (Grant No. FTI_24_2016 of the Tomsk Polytechnic University). T.A. Ishkhanyan acknowledges the support from SPIE through a 2017 Optics and Photonics Education Scholarship, and thanks the French Embassy in Armenia for a doctoral grant, as well as the Agence universitaire de la Francophonie for a joint with the State Committee of Science of Ministry of Education of Armenia Scientific Mobility grant.